# Physics of Bamboo Rifle

Ardi Khalifah

*SMA PGII 1, Bandung, Jl. Panatayuda 2, Bandung 40132, Indonesia*

Mikrajuddin Abdullah

*Center for Interdisciplinary Physics, Department of Physics, Bandung Institute of Technology, Jl. Ganesa 10, Bandung 40132, Indonesia*

*mikrajuddin@gmail.com*

We conducted simple experiments and derived physics equations to explain the working principle of a bamboo rifle. This toy is often played by children in many places that have bamboo plants, such as in ASEAN countries. We identified two main physical processes: mechanics and thermodynamics of gas which controls the firing mechanism of such a rifle.

*Keywords*: Bamboo rifle; experiment; modeling; friction force; adiabatic.

## 1. Introduction

Daily phenomena around us or traditional games can become interesting educational research topics.[1-10] Some of that games may provide amusing ways to teach physics to a range of audiences.[1] The bamboo rifle is an example of a game tool that is played happily by children in various countries that have bamboo plants (**Fig. 1**). It is made of one segment of a bamboo branch, like a hollow pipe, and a slightly shorter bamboo stick to push the bullet through the column. The rear end of the stick is fixed on a handle. Various wet materials can be used as bullets: fruit flesh, soft grains, and the most common, wet papers. The bullet is ejected from the pipe's front end when pushing the stick.

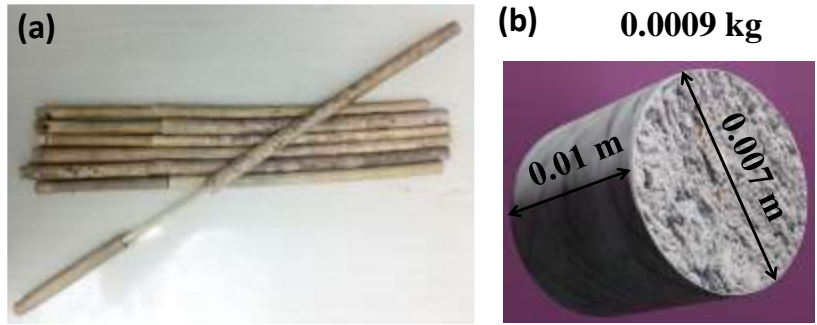

**Fig. 1.** (a) Bambbo rifle toys used in the experiment. (b) Illustration of the bullet when it is in the bamboo column.

However, there are rare reports of the working principle of the bamboo rifle. In this paper, we discuss the physics underlying the working principle of this toy. Simple experiments were also conducted to test the model.

## 2. Experimental Investigation

### 2.1 Bamboo rifle working processes

The working processes of the bamboo rifle are as follows.

1. The first bullet is inserted at the rear end of the pipe and pushed with the stick to occupy the pipe's front end.
2. The second bullet is inserted at the pipe's rear end to trap air in the column (**Fig. 2(a)**).
3. The second bullet is then pushed rapidly with the stick. The volume of column air decreases and the pressure, $P$, increases. The trapped air pushes the first bullet with a force $F = PA$, where $A$ is the cross-sectional area of the bullet (**Fig. 2(b)**).
4. At the same time, the front bullet receives a static frictional force from the column's inner wall and a force by the outer air of $F_o = P_o A$, where $P_o$ is the atmospheric pressure (**Fig. 2(c)**).
5. If the maximum static frictional force is $f_s$, the maximum resisting force acting on the front bullet is $f_s + F_o$. The bullet starts moving if $F > f_s + F_o$.

6. After the bullet moves (**Fig. 2(d)**), the frictional force between the bullet and the pipe's wall changes to the kinetic frictional force $f_k$ so that the total resisting force on the bullet becomes $f_k + F_o$. The acceleration of the bullet is $a = (F - f_k - F_o)/m = (\Delta P A - f_k)/m$ where $\Delta P = P - P_o$ and $m$ is the bullet mass.

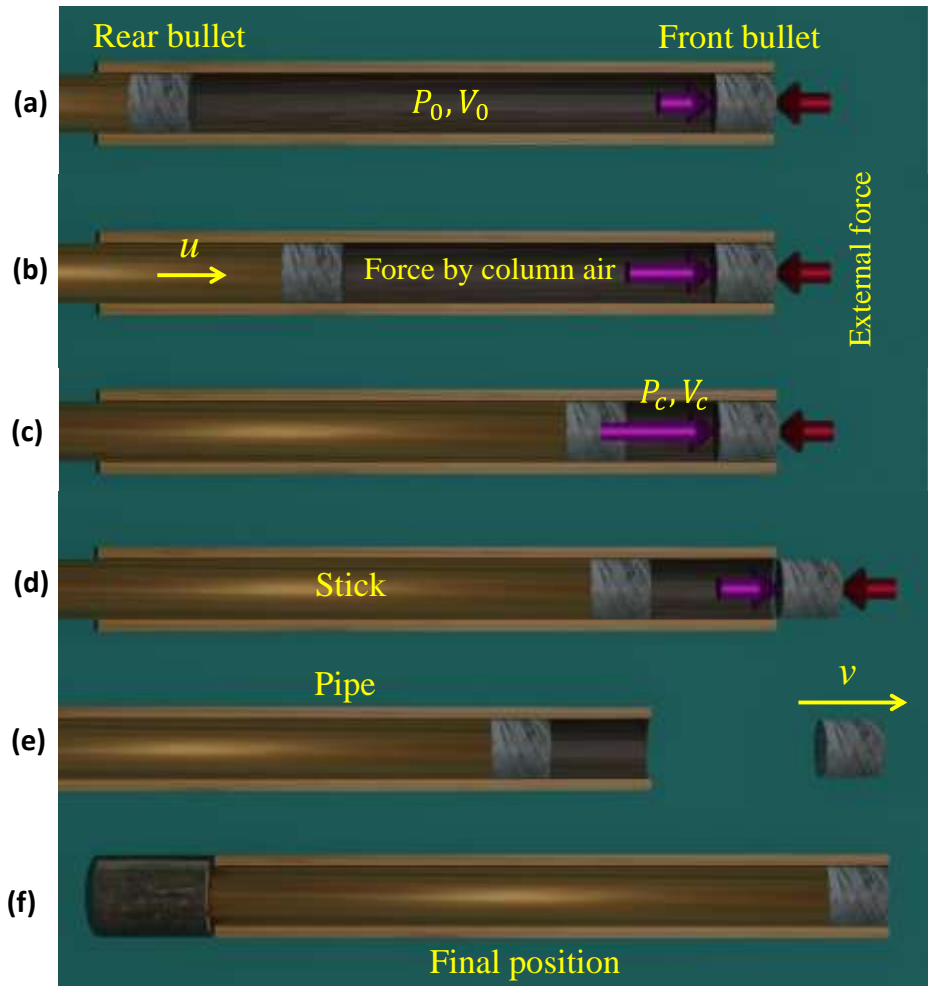

**Fig. 2** Illustration of the shooting process with the bamboo rifle. (a) A new bullet is inserted at the rear end, (b) the hand pushes the new bullet using the stick, (c) critical volume and pressure are reached, (d) the front bullet is starting to leave the rifle, (e) the front bullet is flying at a considerable speed, and (f) the second bullet occupies the position originally occupied by the first bullet. The arrows represent the forces acting on the two ends of the front bullet generated by the air.

## 2.2 Experiment 1: Measuring Bullet Ranges and Ejection Speeds

After leaving the rifle, the bullet moves with a horizontal velocity component much greater than the vertical one. In the horizontal direction, the bullet receives a drag force from the air, $f_d = (1/2)C_d \rho A v_x^2$ where $C_d$ is the drag coefficient, $\rho$ is the density of the air, and $v$ is the speed relative to the air. For short cylindrical objects, $C_d \approx 1.15$.[11] We ignore the drag force in the vertical direction because the velocity component is very small. The components of velocity satisfy the equations, $m dv_x/dt = -(1/2)C_d \rho A v_x^2$ and $v_y = -gt$. The solution for $v_x$ is $v_x(t) = v_0/(1 + b v_0 t)$ where $b = C_d \rho A / 2m$. The positions at any time are $y(t) = H - (1/2)gt^2$ and $x(t) = \ln(1 + b v_0 t)/b$. The bullet hits the ground at $T = \sqrt{2H/g}$ so that the horizontal range of the bullet is $R = x(T) = \ln(1 + b v_0 T)/b$. The bullet ejection speed is

$$v_0 = \frac{1}{bT}(\exp(bR) - 1) \qquad (1)$$

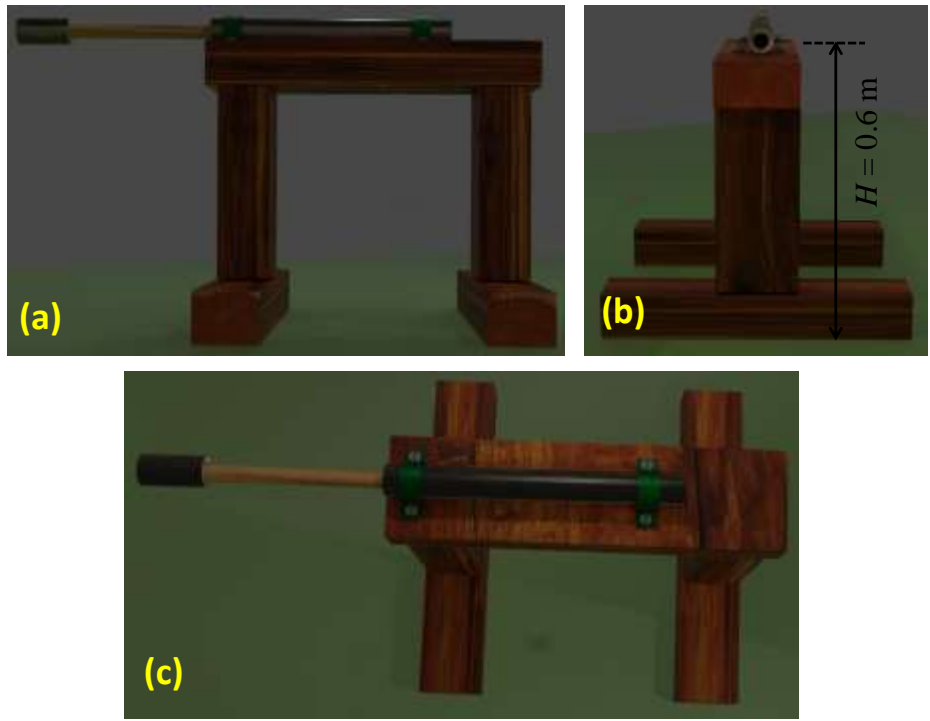

**Fig. 3** The design of the bullet range measuring tool: (a) side view, (b) front view, and (c) top view.

We used a bamboo rifle purchased from an online shop with a column length of 0.3 m and a front column diameter of 0.007 m and an acrylic rifle of the same length but a column diameter of 0.008 m. The bullet mass is about 0.0009 g. The rifle is fired horizontally from a height of 0.6 m (**Fig. 3(a)-(c)**) so that the bullet's flight time is $T = \sqrt{2 \times 0.6/9.8} = 0.35$ s. Using $\rho = 1$ kg/m³, we get $b = 0.025$ m⁻¹. It is clear from Eq. (1), if the air drag is negligible, $b \to 0$ and $\exp(bR) \to 1 + bR$ so that $v_0 \to R/T$, as expected.

From 22 times of firing, we obtained the average ranges as shown in **Table 1**. The greatest range was achieved by bullets made of photocopy paper, then bullets made of cardboard, and lastly bullets made of newsprint. A rifle made of the acrylic pipe using newsprint bullets produced a shorter range. The average calculated ejection speeds are also shown in **Table 1**.

**Table 1** Average ejection speed of the bullet and the average pushing force for bamboo and acrylic rifles using various materials for bullets (distance $D = 0.01$ m). The data is an average of 22 shots.

| Rifle material | Bullet material | Average range (standard deviation) (m) | Average initial speed (standard deviation) (m/s) | Average acceleration (standard deviation) ($\times 10^3$ m/s²) |
|---|---|---|---|---|
| Bamboo | Newsprint | 12.9 (1.6) | 43.5 (6.2) | 96.3 (26.7) |
| | Photocopy paper | 13.7 (1.6) | 46.7 (6.2) | 111.1 (29.1) |
| | Cardboard paper | 13.1 (1.4) | 44.3 (5.6) | 99.6 (24.9) |
| Acrylic | Newsprint | 5.1 (0.6) | 15.6 (2.1) | 12.4 (3.2) |

## 2.3 Experiment 2: Estimating the Friction Force of the Bullet and the Rifle

The front and rear ends of the pipe are occupied by bullets with a length of about 0.01 m each, leaving the initial length of the air column of about 0.28 m. The initial pressure of air in the column is the atmospheric pressure. We observed the bullet is ejected when the column length was reduced to about 0.04 - 0.05 m. Let us take the mean value of 0.045 m. Assuming the air compression process is quasistatic adiabatic, the pressure when the bullet is ejected, $P_c$, satisfies the equation $P_c V_c^\gamma = P_o V_o^\gamma$ where $V_c \cong 0.045A$, $V_o = 0.28A$, and $\gamma = C_p/C_v$ (the ratio of the heat capacity of gas at constant pressure and at constant volume). Using $\gamma = 5/3$, one obtains

$$P_c \cong P_o \frac{0.28^{\frac{5}{3}}}{0.045^{\frac{5}{3}}} \cong 21 P_o \cong 21 \times 10^5 \text{ Pa}$$

This pressure is required to overcome the maximum resistant force acting at the bullet. The cross-sectional area of the bullet is $A = \pi r^2 = \pi(0.007/2)^2 = 3.8 \times 10^{-5}$ m² so that $P_c A \approx 80$ N. The maximum static friction force between the bullet and the pipe wall is about $f_s = P_c A - P_o A = 80 - 3.8 \cong 76.2$ N. **Table 2** shows the average calculated maximum static frictional force on the bullet.

After the bullet moves, the frictional force changes to the kinetic one, $f_k = P'A - P_o A - m\langle a \rangle$, where $P'$ is the column pressure when the bullet is moving. Because the stick is constantly pushed, the air pressure in the column might be different from $P_c$, greater or less. Since the bullet displaces only momentarily before leaving the pipe, we can take $P' \approx P_c$ so that $f_k \approx P_c A - P_o A - m\langle a \rangle = 80 - 3.8 - 0.0009 \times 69961 \cong 13.2$ N.

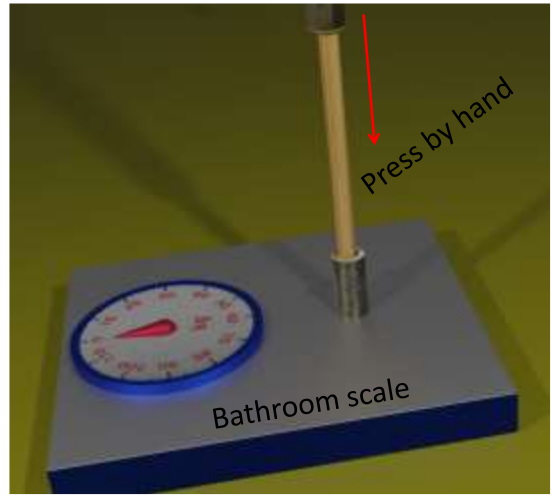

**Fig. 4** The process of measuring the frictional force between the bullet and the bamboo.

For testing this prediction, we conducted an experiment as illustrated in **Fig. 4**. The rifle was gently pressed toward the bathroom scale. The position of the scale needle is recorded with a camera and then played back using the Wondershare Filmora video editor.[12] We obtained the masses shown by the bathroom scale when the bullet starts moving were 6 – 10 kg, 10 -12 kg, and

9 - 13 kg when using bullets made of newsprint, photocopy paper, and cardboard, respectively. The corresponding forces are shown in **Table 2**. Using only the front bullet pushed with the stick we get the frictional force. The maximum friction force between a newsprint bullet and the acrylic pipe with a column diameter of 0.008 m is only about 9.8 N.

To estimate the kinetic friction force, we observed the bathroom scale when the stick pushes the bullet in motion. The scale indication varies between 2 - 3 kg for the three types of paper above, corresponds to forces between 19.6 – 29.4 N, which is assumed to be the kinetic friction force. The estimated value for bullets from newsprint is about 13.2 N.

**2.4 Why the Friction Force is so Large**

Suppose that the maximum static frictional force and the kinetic frictional force satisfy the equations $f_s = \mu_s N$ and $f_k = \mu_k N$ where $N$ is the normal force around the bullet. The coefficient of static friction for most materials is generally below 1.5,[13] so $N = f_s/\mu_s \approx 76.2/1.5 = 50.8$ N. Using this value, the coefficient of kinetic friction is approximately $\mu_k = f_k/N \approx 13.2/50.8 = 0.26$,[13] a reasonable value. Why is the normal force so large? We argue that it is caused by the nonuniformity of cross-section of the column. If moving from back to front, the bamboo column's cross-section gets smaller. We used a bamboo rifle which had a column diameter of 7 mm at the front and 8.5 mm at the rear.

To further test this hypothesis, we measured the static friction force of the bullet as it begins to move from various positions in the column. A bullet at a distance of 0.05 m from the rear end requires a pushing force of about 49 N to start moving. A bullet at a distance of 0.25 m from the rear end needs a pushing force of about 59 N to start moving. The bullet at the exit (0.29 m) requires a pushing force of about 98 N. So, the closer to the front end of the rifle, the maximum static friction force is greater due to the greater normal force.

**Table 2** Average maximum pushing force to displace the bullet.

| Rifle material | Bullet material | Average calculated maximum pushing force (standard deviation) (N)* | Average measured maximum pushing force (N) |
|---|---|---|---|
| Bamboo | News paper | 86.7 (24.0) | ~ 59 – 98 N |
|  | Photocopy paper | 100.9 (26.2) | ~ 98-118 |
|  | Cardboard paper | 89.6 (22.4) | ~ 88 - 127 |
| Acrylic | News paper | 11.1 (2.9) | ~ 9.8 |

*The average frictional force is reduced by $P_0 A \cong 3.8$ N

The small friction force between the acrylic and the bullet made of the newsprint as in **Table 2** also supports this hypothesis. The acrylic has a uniform cross-section. Although the acrylic surface is slicker than bamboo, so the coefficient of friction is smaller, but is likely not too small. As a comparison, the friction coefficient of acrylic and steel is 0.4 – 0.5 while that of wood and steel is around 0.5-0.6 (comparable). This strongly indicates that what plays a role in increasing the frictional force on the bamboo is the non uniformity of cross-sectional size.

To further strengthen the above hypothesis, we reversed the position of the bamboo rifle by shooting from a small cross-section toward the large cross-section. We observed the bullets are not ejected but just fall at the end of the pipe. This is because the front bullet does not get a significant frictional force. High air pressure is not generated in the column. When the air pressure in the column slightly increases, the front bullet readily displaces.

## 3. Modeling

### 3.1 Derivation of Equations

To construct a model, we assume the cross-section of the bamboo column is uniform. The rate change of the column diameter is only (8.5 mm -7 mm)/300 mm = 0.005 = 0.5% so it can be ignored. The volume of the air column when the front bullet is at rest is

$$V(t) = (L_0 - y)A \tag{2}$$

where $L_0$ is the initial length of the air column and $y$ is the displacement of the second bullet. Since the process is quasistatic adiabatic,

$$P_0 V_0^\gamma = P(t) V(t)^\gamma \qquad (3)$$

The pushing force on the front bullet by the column air is

$$F(t) = P(t) A \qquad (4)$$

The front bullet moves when $F(t) - f_s - P_0 A > 0$. The bullet is about to move when the column volume is equal to the critical one, $V_c$, which fulfills

$$\frac{P_0 V_0^\gamma A}{V_c^\gamma} = f_s + P_0 A \qquad (5)$$

When the front bullet is moving, the net force is $\Delta F = F - (f_k + P_0 A)$, or

$$\Delta F = \frac{P_0 V_0^\gamma A}{V(t)^\gamma} - (f_k + P_0 A) \qquad (6)$$

Derivation of the equations in detail is described in Appendix. Finally, we obtain the bullet ejection speed as

$$\frac{v_o}{\beta u/L_c} = \frac{(1-x_o)^{1-\gamma}}{(\gamma-1)} - \frac{L_c^\gamma}{L_0^\gamma}\left(1 + \frac{f_k}{P_0 A}\right) x_o - \frac{1}{(\gamma-1)} \qquad (7)$$

where $\beta = P_0 L_0^\gamma A L_c^{2-\gamma}/mu^2$ and $x_o$ is the solution of the following equation

$$\frac{D}{\beta} = -\frac{(1-x_o)^{2-\gamma}}{(2-\gamma)(\gamma-1)} - \frac{L_c^\gamma}{2L_0^\gamma}\left(1 + \frac{f_k}{P_0 A}\right) x_o^2 - \frac{x_o}{(\gamma-1)} + \frac{1}{(2-\gamma)(\gamma-1)} \qquad (8)$$

Equation (8) must be solved numerically.

## 3.2 Modeling Results

For the simulation purposes, we used data belong to the bamboo rifle: the bullet diameter of 0.007 m, the column initial length of 0.28 m, and the bullet mass of 0.0009 kg. We also used $f_k \cong 13.2$ N and $P_0 A \cong 3.8$ N. The critical length of the air column was varied 0.04 m, 0.06 m, and 0.08 m.

Thus, the parameters used are close to the real values of the bamboo rifle. In the simulation, we vary the stick speed between 1.0 m/s to 8.0 m/s. With these data, we get $\beta = 518 L_c^{1/3}/u^2$.

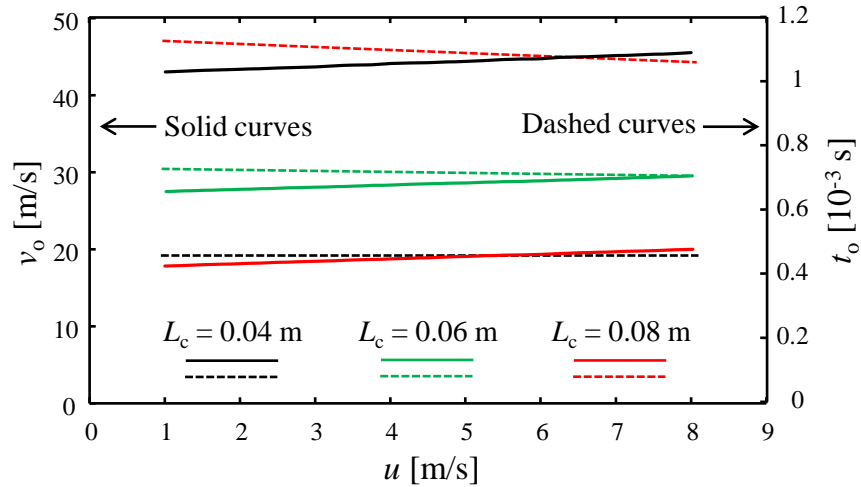

**Fig. 5** Solid curves are the ejection speed of the bullet as a function of the stick speed. Dashed curves are the length of time the bullet moves in the gun. The simulation parameters used are: $L_0 \cong 0.28$ m, $r \cong 0.0035$ m, $P_0 \cong 10^5$ Pa, $m \cong 0.0009$ kg, $D = 0.01$ m, and $\gamma = 5/3$ resulting in $\beta \cong 518 L_c^{1/3}/u^2$. The $L_c$ values are varied: (red) 0.04 m, (blue) 0.06 m, dan (black) 0.08 m.

Solid curves in **Fig. 5** are the dependence of the bullet ejection speed on the stick speed. The smaller the critical length, the higher the bullet's ejection speed. This is due to the formation of greater air pressure in the column when the critical length is smaller. The speed of the stick only a slight affect the ejection speed because the bullet only moves momentarily in the column. As soon as the air pressure exceeds the critical pressure, the bullet immediately leaves the rifle in a period of milliseconds or less. The bullet's ejection speeds in three simulation parameters are between 18 to 46 m/s, close to the measurement results (**Table 1**).

The maximum static frictional force is obtained from Eq. (5). The critical lengths, $L_c = 0.04$ m, 0.06 m, and 0.08 m correspond to the frictional forces of 97.3 N, 49.5 N, and 30.7 N. The duration of the bullet displacement in the rifle is shown by the dashed curves in **Fig. 5**. We find that the order of time is in milliseconds. The shorter the critical length, the smaller the bullet's resident time.

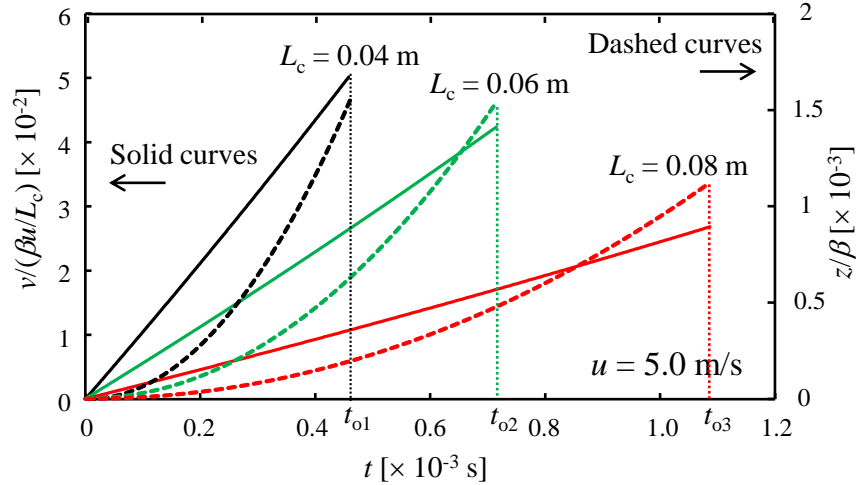

**Fig. 6** Evolution of the front bullet displacement (normalized to $\beta$) and speed (normalized to $\beta u/L_c$). The simulation parameters are the same as those used in determining the data in **Fig. 5** except for those written in the graph.

The evolution of the bullet's position and velocity as a function of $x = ut/L_c$ is shown in **Fig. 6**. We only calculate for three critical lengths: $L_c$ = 0.04 m, 0.06 m, and 0.08 m and one stick speed: $u = 5$ m/s. Other parameters are the same as those used to calculate the data in **Fig. 5**. Solid curves are the ejection speed while dashed curves are the displacement. The shorter the critical length, the shorter the time for the bullet to leave the rifle. By using a bullet length of about 0.01 m, the average acceleration of the bullet is in the order of $2 \times 0.01/(10^{-3})^2 = 2 \times 10^4$ m/s², comparable to the data in **Table 1**.

## 4. Conclusion

We have derived equations to explain the working principle of the bamboo rifle. The equations estimated values that are close to the experimental results (at least in the order of magnitude). This demonstrates that children's games can be a challenging research topic and explainable using basic physics.

**Appendix**

For simplicity, let us take $L = L_c$ and $y = 0$ at $t = 0$ so that when the front bullet is moving, the volume of the air column is $V = (L_c - y + z)A$ where $z$ $(0 \leq z \leq D)$ is the front bullet displacement from its resting position. Furthermore, let us assume $z \ll L_c - y$ so that the volume of the air column is $V(t) \cong (L_c - y)A$. We also assume that the rod speed, $u$, is constant, so we can write $y = ut$. The acceleration of the front bullet is $a = d^2z/dt^2 = \Delta F/m$, or

$$\frac{d^2z}{dt^2} = \frac{P_0 L_0^\gamma A}{mL_c^\gamma(1-ut/L_c)^\gamma} - \frac{f_k + P_0 A}{m} \tag{S1}$$

If we write $x = ut/L_c$, $d^2z/dt^2 = (u^2/L_c^2)d^2w/dx^2$ and Eq. (S1) becomes

$$\frac{d^2z}{dx^2} = \frac{\beta}{(1-x)^\gamma} - \frac{L_c^2(f_k + P_0 A)}{mu^2} \tag{S2}$$

where

$$\beta = \frac{P_0 L_0^\gamma A L_c^{2-\gamma}}{mu^2} \tag{S3}$$

The general solution of Eq. (S2) is

$$z(x) = \frac{\beta}{(\gamma-2)(\gamma-1)}(1-x)^{2-\gamma} - \frac{L_c^2(f_k+P_0A)}{2mu^2}x^2 + c_1 x + c_2$$

where $c_1$ and $c_2$ are constants. The initial condition $z(0) = 0$ results in $c_2 = \beta/(2-\gamma)(\gamma-1)$.

The speed of the front bullet is

$$v = \frac{dz}{dt} = \frac{dz}{dx}\frac{dx}{dt} = \frac{u}{L_c}\left[\frac{\beta}{(\gamma-1)}(1-x)^{1-\gamma} - \frac{L_c^2(f_k+P_0A)}{mu^2}x + c_1\right] \tag{S4}$$

The initial condition $v(0) = 0$ results in $c_1 = -\beta/(\gamma-1)$. Finally, the complete solution for the front bullet position can be written as

$$\frac{z(x)}{\beta} = -\frac{(1-x)^{2-\gamma}}{(2-\gamma)(\gamma-1)} - \frac{L_c^2(f_k+P_0A)}{2\beta mu^2}x^2 - \frac{x}{(\gamma-1)} + \frac{1}{(2-\gamma)(\gamma-1)}$$

$$= -\frac{(1-x)^{2-\gamma}}{(2-\gamma)(\gamma-1)} - \frac{L_c^\gamma}{2L_0^\gamma}\left(1+\frac{f_k}{P_0A}\right)x^2 - \frac{x}{(\gamma-1)} + \frac{1}{(2-\gamma)(\gamma-1)} \tag{S5}$$

and the front bullet speed before leaving the pipe can be written as

$$\frac{v(x)}{\beta u/L_c} = \frac{(1-x)^{1-\gamma}}{(\gamma-1)} - \frac{L_c^\gamma}{L_0^\gamma}\left(1+\frac{f_k}{P_0 A}\right)x - \frac{1}{(\gamma-1)} \qquad (S6)$$

We will determine the speed of the bullet as it leaves the rifle. The bullet leaves the rifle when $z = D$ which occurs at $x_o$. By using Eq. (A5) we get the equation fo $x_o$ as follows,

$$\frac{D}{\beta} = -\frac{(1-x_o)^{2-\gamma}}{(2-\gamma)(\gamma-1)} - \frac{L_c^\gamma}{2L_0^\gamma}\left(1+\frac{f_k}{P_0 A}\right)x_o^2 - \frac{x_o}{(\gamma-1)} + \frac{1}{(2-\gamma)(\gamma-1)} \qquad (S7)$$

and the bullet speed when leaving the front end of the pipe is given by Eq. (7)

$$\frac{v_o}{\beta u/L_c} = \frac{(1-x_o)^{1-\gamma}}{(\gamma-1)} - \frac{L_c^\gamma}{L_0^\gamma}\left(1+\frac{f_k}{P_0 A}\right)x_o - \frac{1}{(\gamma-1)} \qquad (7)$$

## Videos

The illustration videos can be found in the folloding links

https://drive.google.com/file/d/1FUDGuAv_x_GEDpHIHZnhwjs8MhXTA9KV/view?usp=sharing

https://drive.google.com/file/d/1i6ZWn36kG5I2XOPZz5j-Qklh2dGyd3Bo/view?usp=sharing

https://drive.google.com/file/d/1P2zafV8v8gNSDdk9z_IqejpumV6zsiOY/view?usp=sharing